\documentclass{aptpub}
\authornames{D. J. SMITH AND J. R. BLAKE}
\shorttitle{Surface accumulation of spermatozoa}

\usepackage[pdftex]{graphicx}

\usepackage{float}
\usepackage{amsmath}
\newcommand\pd[3][]{\frac{\partial^{#1} #2}{\partial{#3}^{#1}}}
\newcommand{\bs}{\boldsymbol}
\newcommand{\smum}{\, \mu \mathrm{m}}
\newcommand{\mums}{\, \mu \mathrm{m}/\mathrm{s}}

  \renewcommand{\figurename}{\textsc{Figure}}

\begin{document}
\title{Surface accumulation of spermatozoa: a fluid dynamic phenomenon}

\authorone[University of Birmingham]{D. J. Smith}
\addressone{School of Clinical and Experimental Medicine and School of Mathematics, University of Birmingham, Edgbaston, Birmingham, B15 2TT, UK; Centre for Human Reproductive Science, Birmingham Women's NHS Foundation Trust, Metchley Park Road, Edgbaston, B15 2TG, UK}

\authortwo[University of Birmingham]{J. R. Blake}
\addresstwo{School of Mathematics, University of Birmingham, Edgbaston, Birmingham, B15 2TT, UK; Centre for Human Reproductive Science, Birmingham Women's NHS Foundation Trust, Metchley Park Road, Edgbaston, B15 2TG, UK}

\begin{abstract}
Recent mathematical fluid dynamics models have shed light into an outstanding problem in reproductive biology:
why do spermatozoa cells show a `preference' for swimming near to surfaces? In this paper we review quantitative
approaches to the problem, originating with the classic paper of Lord Rothschild in 1963. A recent `boundary integral/slender body theory' 
mathematical model for the fluid dynamics is described, and we discuss how it gives insight
into the mechanisms that may be responsible for the surface accumulation behaviour. We use the simulation model to explore these mechanisms in more detail, and discuss whether simplified models can capture the behaviour of sperm cells. The far-field decay of the fluid flow around the cell is calculated, and compared with a stresslet model.
Finally we present some new findings showing how, despite having a relatively small hydrodynamic drag, the sperm cell `head' has very significant
effects on surface accumulation and trajectory.
\end{abstract}

\keywords{fluid dynamics, Stokes flow, spermatozoa, slender body theory}

\ams{76Z10}{92C05, 92C17}

\section{Introduction}
Approximately one in six couples in the world experiences difficulties conceiving children, and on clinical examination male factors are found to contribute in at least half of all cases. A critical aspect of the function of the human male gamete is its \textit{motility}: the sperm's ability to swim through the liquids of the female reproductive tract to the vicinity of the egg, penetrate the surrounding layers and fertilise. Motility is therefore an important target for novel diagnostic and therapeutic techniques, and advances in this area would have considerable financial and quality-of-life implications in human fertility, domestic animal breeding and wildlife conservation. The ability of sperm to swim is perhaps their most obvious function to the physical scientist, having long-provided an important archetype in mathematical fluid dynamics modelling. In this article, we discuss an important and surprisingly elusive aspect of the swimming behaviour of sperm both \textit{in vivo} and \textit{in vitro}---the phenomenon of `surface accumulation'.

\begin{figure}
$
\begin{array}{ll}
\mbox{(a)}  \hspace{-0.1cm} & \mbox{(b)} \vspace{0.2cm}\\ 
\scalebox{1.0}{
\includegraphics{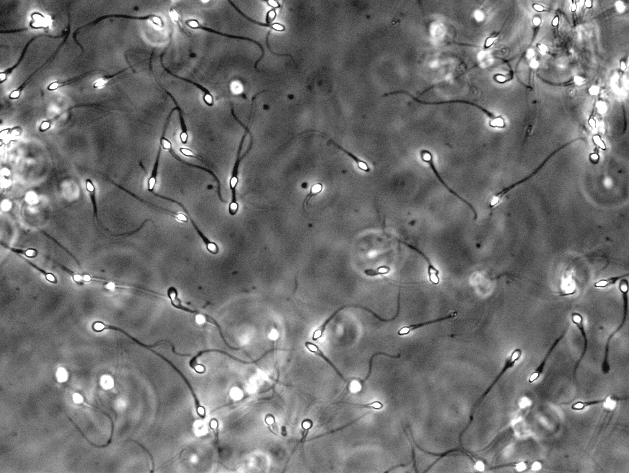}
} \hspace{-0.1cm} &  \scalebox{1.0}{
\includegraphics{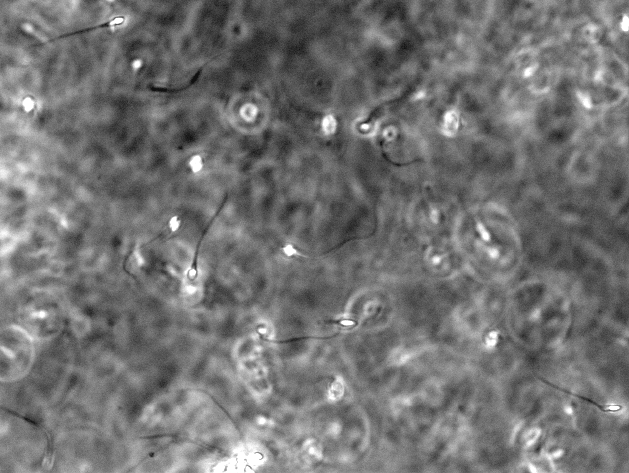}
}
\end{array}
$
\caption{Micrographs showing (a) high sperm cell concentration near a microscope slide, (b) low sperm cell concentration at a level mid-way between the microscope slide and coverslip. \textit{Unpublished data}, Smith and Kirkman-Brown 2009.} \label{baccSchema}
\end{figure}

Rothschild \cite{rothschild} famously observed \textit{the non-uniform distribution of sperm in a drop of bull semen}: spermatozoa typically exhibit a `preference' for surfaces, with by far
the greatest concentrations being found near to the inner faces of imaging chambers or the microscope slide and coverslip (Figure~\ref{baccSchema}). Moreover, once in the vicinity
of a surface, sperm cells tend to remain at a constant distance, a fact that is very convenient in microscopical imaging of the sperm and its flagellar movement, Smith et al.\ \cite{smithcellmo}.
In the sperm of humans and the majority of other species studied, gravity has not been observed to have a significant effect, so that the number of cells accumulating on upper 
and lower surfaces is typically similar, Winet et al.\ \cite{winet84}. A related phenomenon occurs in other motile microscopic cells such as bacteria, and this has long interested researchers, particularly at the interface of mathematical modelling and experimental biology (see \cite{berke08}, \cite{cisneros07}, \cite{laugaecoli}, \cite{winet84}). 

A recent study of surface accumulation of sperm by Smith et al.\ \cite{smithjfm} predicts that, at least for the human gamete, the effect can occur through fluid mechanics alone, resulting from the viscous-dominated regime of microscopic swimming, without the need for a specialised `bias' of the flagellar beat to push the cell towards the surface. In this paper we review the findings of this study, then discuss how the model gives insight into the mechanisms that may and may not be responsible for this behaviour, and in particular some important and previously unexplored aspects that a fundamental theory of surface accumulation must address.

\section{Mathematical modelling of cell swimming at zero Reynolds number}
Fluid flow is modelled mathematically by the Navier-Stokes partial differential equations. For a constant viscosity Newtonian incompressible fluid, these take the following form:
\begin{equation}
\rho \left( \frac{\partial \bs{u}}{\partial t} + (\bs{u} \cdot \bs{\nabla}) \bs{u} \right) = - \bs{\nabla} p + \mu \nabla^2 \bs{u} \mbox{,} \quad \bs{\nabla} \cdot \bs{u}  =  0.
\end{equation}
The dependent variables are the velocity field $\bs{u}=\bs{u}(\bs{x},t)$ representing the fluid velocity at location $\bs{x}$ and time $t$, and the pressure field $p=p(\bs{x},t)$. The constants $\rho$ and $\mu$ are density and viscosity, the latter quantity being measured in \textit{Pascal seconds} (Pa s) in SI units. The first equation is derived from the principle
of momentum balance, with the terms on the left hand side arising from inertia, while the terms on the right hand side arise from pressure and viscosity respectively. The second equation corresponds to the principle of mass conservation, combined with incompressibility.

The most famous source of mathematical difficulty is the nonlinear term $(\bs{u} \cdot \bs{\nabla}) \bs{u}$, from which arises the great unsolved problem of turbulence and the open question of whether the Navier-Stokes equations always have smooth solutions. However, in sperm fluid mechanics it is possible to neglect this term: nondimensionalising the equations with $U$ as velocity scale, $L$ as length scale, $L/U$ as time scale and $\mu U/L^2$ as pressure scale, the momentum equation is simplified to
\begin{equation}
\frac{\rho U L}{\mu} \left( \frac{\partial \bs{u}}{\partial t} + (\bs{u} \cdot \bs{\nabla}) \bs{u} \right)  =  - \bs{\nabla} p + \nabla^2 \bs{u} \mbox{,} \label{ndns}
\end{equation}
where the coefficient on the left hand side is the Reynolds number $R=\rho U L/\mu$, first identified by Osborne Reynolds as the parameter governing the transition to turbulence. A sperm cell has length approximately $50 \smum$, swims with velocity approximately $50 \mums$, and the density and viscosity of saline solution can be estimated as $10^3 \; \mathrm{kg}/\mathrm{m}^3$ and $10^{-3} \; \mathrm{Pa} \; \mathrm{s}$. We estimate the Reynolds number as $2.5 \times 10^{-3}$, and indeed in solutions of higher viscosity it will be smaller still. It is therefore an excellent approximation to consider the `zero Reynolds number' \textit{Stokes flow} equations,
\begin{equation}
0 = - \bs{\nabla} p + \nabla^2 \bs{u} \mbox{,} \quad \bs{\nabla} \cdot \bs{u} = 0 \mbox{.} \label{stokes}
\end{equation}
Solving this equation directly using the computational fluid dynamics approaches generally applied to higher Reynolds number flows is computationally expensive, since there are three spatial dimensions. Simulating the trajectory of a sperm additionally involves solving the equation at a sequence of timesteps, and an additional complication is the presence of moving boundaries due to the movement of the flagellum and head, which in a computational fluid dynamics solution would necessitate moving the computational mesh at every timestep, which incurs further expense.
The analysis we use in this paper exploits the linearity of equation~\eqref{stokes} to solve the problem much more efficiently.

Very low Reynolds number corresponds to the near-absence of inertia, and in this respect, the fluid mechanics of sperm swimming differs significantly from more familiar situations such as human or fish swimming, or indeed most of the fluid dynamic phenomena we encounter. An important property of equation~\eqref{stokes} is the absence of explicit time dependence, leading to \textit{time-reversal symmetry}. Classic demonstrations of this were given by Sir Geoffrey Taylor in his 1967 film \textit{Low Reynolds number flow}, including the `smearing' and `unsmearing' of an ink droplet in glycerine, and the abolition of swimming of a model fish by high viscosity liquid. This latter effect is a crucial aspect of sperm swimming---only by performing a time-irreversible motion, such as propagating a bending wave, can sperm swim effectively. `Flapping' motions of a fish tail or other reciprocating movements are ineffective---later referred to by Edward Purcell \cite{purcell77} as the `Scallop Theorem'. Recent research by Lauga \cite{laugascallop} has shown that in non-Newtonian liquids, such as polymer suspensions, the introduction of an explicit time-dependence results in the breakdown of this theorem.

\section{Singular solutions of the Stokes flow equations}
The problem of sperm swimming will be formulated using equation~\eqref{stokes}, together with boundary conditions for the fluid velocity on the surface of the sperm tail, head, and additionally a nearby glass surface. The linearity of equation~\eqref{stokes} allows the superposition of solutions in order to satisfy the boundary conditions, and the techniques we shall use will be based on singular solutions, in particular the Stokeslet (point-force), Stokes-dipole (point-force dipole) and the potential flow source dipole.

The Stokeslet $S_{jk}(\bs{x},\bs{y})$ is defined as the solution $\bs{u}(\bs{x})=(S_{1k}(\bs{x},\bs{y})$, $S_{2k}(\bs{x},\bs{y})$, $S_{3k}(\bs{x},\bs{y}))$ and $p(\bs{x})=P_k(\bs{x},\bs{y})$ of
\begin{equation}
-\bs{\nabla} p + \nabla^2 \bs{u} + \bs{e}_k\delta(\bs{x}-\bs{y})  = 0 \mbox{,} \quad \bs{\nabla} \cdot \bs{u} = 0 \mbox{,}  \label{stokes2}
\end{equation}
which is the Stokes flow equation with a concentrated unit force per unit volume located at $\bs{y}$, pointing in the $k$-direction, with $\delta(\bs{x}-\bs{y})$ denoting the Dirac delta distribution.  The solution for a point force in an infinite fluid is given by $S_{jk}=(\delta_{jk}/r+r_jr_k/r^3)/8\pi$ and $P_k=r_j/(4\pi r^3)$, where $r_j=x_j-y_j$ and $r^2=r_1^2+r_2^2+r_3^2$, and $\delta_{jk}$ denotes the Kronecker delta tensor. 
Other singular solutions include the Stokes-dipole, defined as $S^D_{jkl}=(\partial/\partial x_l) S_{jk}$, and the potential source-dipole $P^D_{jk}=(\partial/\partial x_k)(r_j/r^3))/4\pi$. The latter is named due to being a source-dipole solution of the potential flow equations, and is a solution of equation~\eqref{stokes} with zero pressure field. 

The \textit{Stokeslet}, coined by Hancock \cite{hancock} as a name for the \textit{purely viscous component of the Stokes flow around a translating sphere}, the other component being the potential source-dipole. The Stokeslet and potential-dipole tensors are singular at the centre of the sphere, so that the flow is regular in the domain $|\bs{x}-\bs{y}|>a$, where $a$ is the sphere radius. The concept of representing flows around finite bodies by combinations of singularities outside or on the boundary of the flow domain forms the basis for the slender body theory and boundary integral method we discuss below.

\section{The no-slip condition in viscous fluid mechanics and the method of images}
The appropriate velocity boundary condition in viscous flow is the no-slip, no-penetration condition $\bs{u}(\bs{x}=\bs{X})=\bs{U}(\bs{X})$, where $\bs{X}$ is the position vector of a solid surface and $\bs{U}(\bs{X})$ is its velocity. In this problem we are concerned with a stationary boundary representing the glass surface of a microscope slide or a coverslip, and a moving boundary representing the cell head and tail surfaces. The glass surface is modelled as the plane $x_3=0$, with flow taking place in the `half-space' region $x_3>0$. In the experiment, there is an additional bounding surface, but since we shall only consider motion close to one surface, this is neglected in our model. Liron and Mochon \cite{liron76s} showed how to model the effect of the additional surface, but this is more computationally expensive, as discussed in Smith et al.\ \cite{smithdsl}.

Blake \cite{bl71sto} addressed the problem of modelling a flow using Stokeslet distributions while satisfying the no-slip condition $\bs{u}(x_3=0)$, motivated by the propulsion of fluid by cilia protruding from a cell surface. The technique is equally useful in modelling the interaction of a glass surface and a sperm cell. We denote the solution here as $\bs{\mathsf{B}}$, where
\begin{eqnarray}
B_{ij}(\bs{x};\bs{\xi}) & = & \frac{1}{8\pi}\left(\left[\frac{\delta_{ij}}{r} + \frac{r_i r_j}{r^3} \right]  		                  
       - \left[\frac{\delta_{ij}}{R} + \frac{R_i R_j}{R^3} \right]             \right. \nonumber \\
& &   \left.         + 2 \xi_3 ( \delta_{j\alpha} \delta_{\alpha k} - \delta_{j3} \delta_{3k})  \pd{}{R_k} 
\left[ \frac{\xi_3 R_i}{R^3} -   \left[\frac{\delta_{i3}}{R}+\frac{R_i R_3}{R^3} \right] \right] \right) \mbox{,} \label{blimage}
\end{eqnarray}
with $R_\alpha=r_\alpha$, $R_3=x_3+\xi_3$ and the index $\alpha$ taking values $1$, $2$ only. Equation~\eqref{blimage}
consists of a Stokeslet in the fluid, an equal and opposite image Stokeslet, and further images consisting of a Stokes-dipole and potential dipole. The tensor $( \delta_{j\alpha} \delta_{\alpha k} - \delta_{j3} \delta_{3k})$ takes the value $+1$ for $j=k=1,2$, value $-1$ for $j=k=3$ and zero otherwise. An equivalent solution had technically been found previously in the 1896 work of Lorentz (see \cite{lorentz}); however it was not until solution~\eqref{blimage} was discovered that the nature of the image systems was clear. Furthermore this result showed how the far-field is modified by the boundary effect: the $O(r^{-1})$ decay of a force acting parallel to the boundary is modified to become a $O(r^{-2})$ decay, the far-field being of `stresslet' character, i.e. a symmetric Stokes-dipole. This has recently been used to understand the fluid dynamics of the movement of vesicles in the developing embryonic node, Smith et al.\ \cite{smithnodal}.

\section{Slender body theory and the boundary integral method} 
Performing the first detailed mathematical analysis of microorganism swimming, G. J. Hancock \cite{hancock}, working with Sir James Lighthill, formulated a slender body theory for Stokes flow, inspired by slender body theory for potential flow. This involved representing the fluid flow around the flagellum using a line integral of Stokeslets weighted by unknown force density. Considerable work on simplifications to slender body theory (see \cite{grayhan}, \cite{lighthill76}), adaptations for cilia modelling (see \cite{bl72}, \cite{fulb}, \cite{gueron92}, \cite{liron76}), refinements to incorporate a spherical cell body (see \cite{higdon79}), large-amplitude motions (see \cite{dresdner}, \cite{higdon79}), asymptotic analysis (see \cite{batchelor70}, \cite{cox70}) and higher-order corrections (see \cite{johnson}) have been presented, as reviewed in detail in by Smith et al.\ \cite{smithjfm}. Essentially, all of these theories involve representing the flow field due to the moving tail by a line integral
\begin{equation}
\bs{u}_{\mathrm{tail}} = \int_0^1 \bs{\mathsf{G}}(\bs{x},\bs{\xi}) \cdot \bs{f}(s) \; \mathrm{d}s \mbox{,}
\end{equation}
where the kernel $\bs{\mathsf{G}}$ is a force singularity Green's function, coupled with a potential-dipole $\bs{\mathsf{P}}^D$ to improve the accuracy of the slender body representation, and $\bs{f}(s)$ is the force per unit length.

In order to model the effect of the sperm cell body or `head', there are three fluid dynamic effects that must be taken into account. In order of importance, they are:
\begin{enumerate}
\item The drag and torque on the translating and rotating head, balancing the force and torque produced by the flagellum,
\item The near and far-fields of the flow generated by the head as it moves through the fluid,
\item The fact that the no-slip condition must be satisfied on the head, taking into account the flow produced by the flagellum. This modifies $\bs{f}(s)$ and usually reduces the swimming speed.
\end{enumerate}
The only fluid dynamic model to have taken all these effects into account while using slender body theory exclusively was that of Higdon \cite{higdon79}, which modelled the swimming of a spherical-headed microorganism. Taking into account the non-spherical head of most sperm cells, and additionally the effect of a nearby no-slip surface, requires the use of more complex numerical implementations, such as the boundary integral method, as first applied to sperm by Phan-Thien et al.\ \cite{phan}. The boundary integral method uses distributions of Stokeslets over the a surface $\partial H$, with force per unit area given by the unknown traction $\phi(\bs{X}^h,t)$ for 
$\bs{X}^h \in \partial H$. Recently, we used a hybrid of the boundary integral method for accurate modelling of the head, and slender body theory for the flagellum (see \cite{smithjfm}), in order to provide an efficient and accurate representation of both parts of the cell. In this method, the velocity field for points in the fluid $\bs{x}$ outside of, or on, the surface of the sperm is given by
\begin{eqnarray}
\bs{u}(\bs{x},t) &=& \bs{u}_{\mathrm{head}}(\bs{x},t) + \bs{u}_{\mathrm{tail}}(\bs{x},t) \nonumber \\
&=& \int \! \! \int_{\partial H} \bs{\mathsf{B}}(\bs{x},\bs{X}^h) \cdot \bs{\phi}(\bs{X}^h,t) \; 
\mathrm{d} \bs{X}^h + \int_0^1 \bs{\mathsf{G}}(\bs{x},\bs{\xi}(s,t)) \cdot \bs{f}(s,t) \; \mathrm{d} s \mbox{.}   \label{fluid_velocity}
\end{eqnarray}
Since the movement of the flagellum will be specified relative to the cell, not a stationary observer, a coordinate transformation is required. The position of the flagellar centreline in the laboratory frame is given by
\begin{equation}
\bs{\xi}	 = \bs{X}_0 + \bs{\mathsf{M}} \cdot \bs{\xi}'\mbox{,}
\end{equation}
where $\bs{X}_0$ is the location of the head/flagellum junction, $\bs{\mathsf{M}}$ is a rotation matrix from the `body frame', in which the head is stationary, to a fixed `lab frame', and $\bs{\xi}'$ is the flagellum position vector in the body frame. The flagellar movement is specified as the function $\bs{\xi}'$.
The body frame will translate and rotate with \textit{a priori} unknown velocity $\bs{V}$ and angular velocity $\bs{\Omega}$. By rigid body mechanics, 
with respect to the laboratory, the kinematic velocities of points on the head $\bs{U}(\bs{X}^h)$ and flagellum 
$\bs{U}(\bs{\xi})$ are respectively
\begin{eqnarray}
\bs{U}(\bs{X}^h) & =  & \mathbf{V} + \bs{\Omega} \wedge (\bs{X}^h - \bs{X}_0) , \nonumber \\
\bs{U}(\bs{\xi})     & =  & \mathbf{V} + \bs{\Omega} \wedge (\bs{\xi} - \bs{X}_0) + \bs{\mathsf{M}} \cdot \dot{\bs{\xi}}' \mbox{,}
\end{eqnarray}
In order to close the system so that $\bs{\phi}$, $\bs{f}$, $\bs{V}$ and $\bs{\Omega}$ can be determined simultaneously, we use the principles of force and torque balance in zero Reynolds number flow:
\begin{eqnarray}
0 &=& \int \!\int_{\partial H} \bs{\phi}(\bs{X}^{h},t) \; \mathrm{d} \bs{X}^{h} 
+ \int_0^1 \bs{f}(s,t) \; \mathrm{d} s \mbox{,} \nonumber \\
0 &=& \int \!\int_{\partial H} (\bs{X}^{h}-\bs{X}_0) \wedge \bs{\phi}(\bs{X}^{h},t) \; \mathrm{d} \bs{X}^{h}  + \int_0^1  
(\bs{\xi}(s,t)-\bs{X}_0) \wedge \bs{f}(s,t) \; \mathrm{d} s \mbox{.} \nonumber \\
& & \label{balance}
\end{eqnarray}

At each instant in time, the initial position and orientation are known, and the flagellum position $\bs{\xi}$, and kinematics $\dot{\bs{\xi}}$, with respect to the head are given by a mathematical formula chosen to represent the flagellum waveform. From this, equations~\eqref{fluid_velocity} and \eqref{balance} form a closed system for the unknowns $\bs{\phi}$, $\bs{f}$, $\bs{V}$ and $\bs{\Omega}$. A collocation method is used to discretise these equations spatially, and a matrix system with typically $3\times(60+32)+3+3$ scalar unknowns is formulated, corresponding to 60 constant-force segments for the flagellum, 32 constant-force elements for the head, the additional terms being the 3 scalar components of $\bs{V}$ and $\bs{\Omega}$. The solution of this system allows the updating of the cell position and orientation from $\bs{V}$ and $\bs{\Omega}$. From the new values of $\bs{f}$ and $\bs{\phi}$, the equation~\eqref{fluid_velocity} can be used to calculate the velocity of the fluid at any point in space. 

This method is very computationally efficient, allowing the accurate simulation of 8000 beat cycles of migration, discretised with 100 timesteps per beat, in approximately 20 hours of CPU time. Such efficient computation of a moving-boundary, 3D, unbounded and time-dependent problem is not possible with standard computational fluid dynamics methods that involve the determination of the fluid velocity at every point in the domain at every timestep. The simplification arises from the linearity of the Stokes flow equations, the corresponding ability to take integral sums of known solutions, and our knowledge of  Green's functions such as $\bs{\mathsf{B}}$. 

Simulations were performed with the University of Birmingham's BlueBear cluster, using a single core per simulation. This allows parameter space to be searched through the simultaneous execution of many simulations; however a single simulation could be performed in a similar time period on a desktop PC.

\section{Prescribing the flagellar beat}
\begin{figure}[h]
\[
\begin{array}{ll}
\mbox{(a), }k=2\pi & \mbox{(b), }k=3\pi \\ 
\includegraphics[clip=true, viewport=.0in .0in 1in 1in]{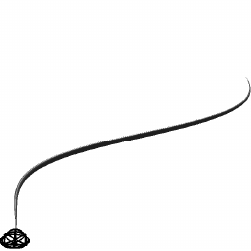} 
& 
\includegraphics[clip=true, viewport=.0in .0in 1in 1in]{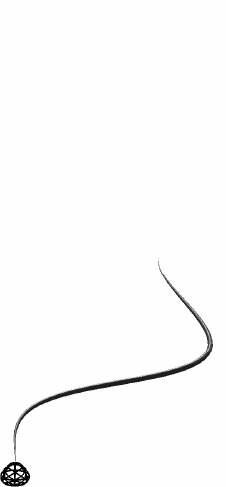}
\end{array}
\]
\caption{Planar sinusoidal waveforms used for the simulations, with (a) $k=2\pi$ and (b) $k=3\pi$. }\label{spermSnapk}
\end{figure}

The remaining step in the model is to specify the movement of the flagellum $\bs{\xi}'$. While some approaches have generated the flagellar beat from a complex interaction of internal mechanical activity and external fluid forces, we have taken the simpler approach of writing $\bs{\xi}'$ as an analytic function. The flagellar waveform of human sperm is usually a three-dimensional shape, likely to resemble a nearly-planar flattened helix (see \cite{ishijima86}). In \cite{smithjfm} we examined the effect of both a planar waveform and a flattened helix, and found that the difference in surface accumulation in the two cases was minimal. Hence in what follows we shall focus on the planar beat, which simplifies the analysis of the flow field around the cell. The waveform we choose was first suggested by Dresdner and Katz \cite{dreskatz81}, and is proportional to a sine-wave $\sin(kx_1-\omega t)$, modified to have linearly increasing amplitude along the flagellum. In \cite{smithjfm}, we found that the angular wavenumber parameter $k$ is the critical parameter governing surface accumulation, as discussed in the next section. Figure~\ref{spermSnapk} shows the waveform for $k=2\pi$ and $k=3\pi$ respectively.

\section{Recent findings}
\begin{figure}[h]
\begin{center}
(a)\\
\scalebox{0.83}{
\includegraphics{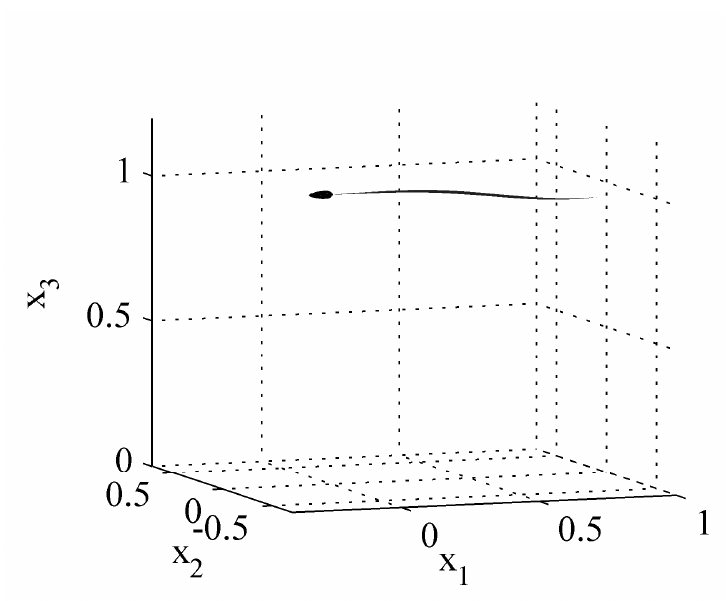}
} \\
(b)\\
\includegraphics{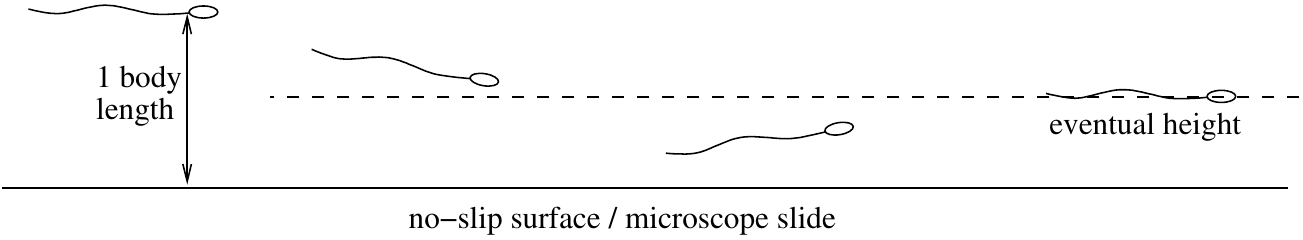}
\end{center}
\caption{(a) The initial configuration for all simulations, with the head/flagellum junction initially at $\bs{x}=(0,0,1)$.
(b) Schematic showing the surface accumulation behaviour predicted by simulations for wavenumbers $k>2.4\pi$. The horizontal distance travelled is an order of magnitude greater than shown in the schematic, as shown in Figure~\ref{trajectories}.}\label{initial}
\end{figure}

Our recent simulation study \cite{smithjfm} mainly concerned the behaviour of a cell initially at $x_3=1$ in non-dimensional units, i.e.\ one body length away from a no-slip boundary at $x_3=0$, with initially parallel orientation, and its beat plane parallel to the no-slip boundary, as shown in the simulation snapshot Figure~\ref{initial}(a).
The behaviour predicted by the simulations for wavenumbers $k>2.4\pi$ is shown schematically in Figure~\ref{initial}(b): the cell initially pitches towards the surface and swims towards it, then pitches away, repeating this behaviour until it approaches an `equilibrium height' at which it remains.

\begin{figure}[H]
(a)\\
\scalebox{0.83}{
\includegraphics{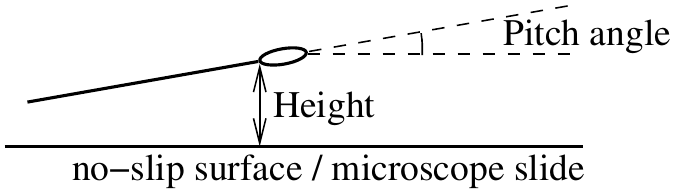}
}\\
(b)\\
\scalebox{0.83}{
\includegraphics{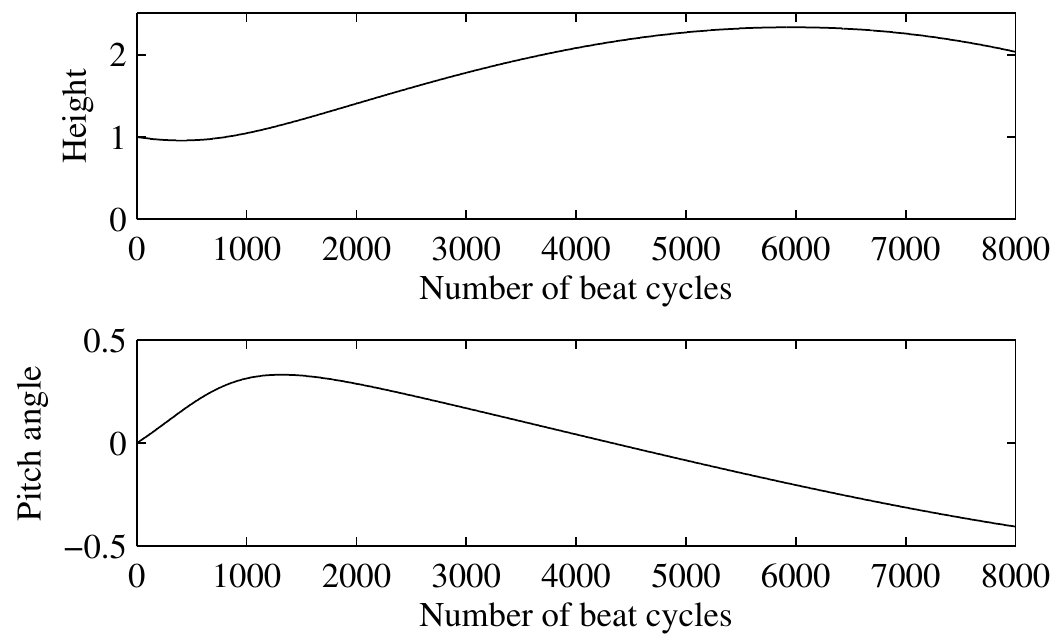}
}\\
(c)\\
\scalebox{0.83}{
\includegraphics{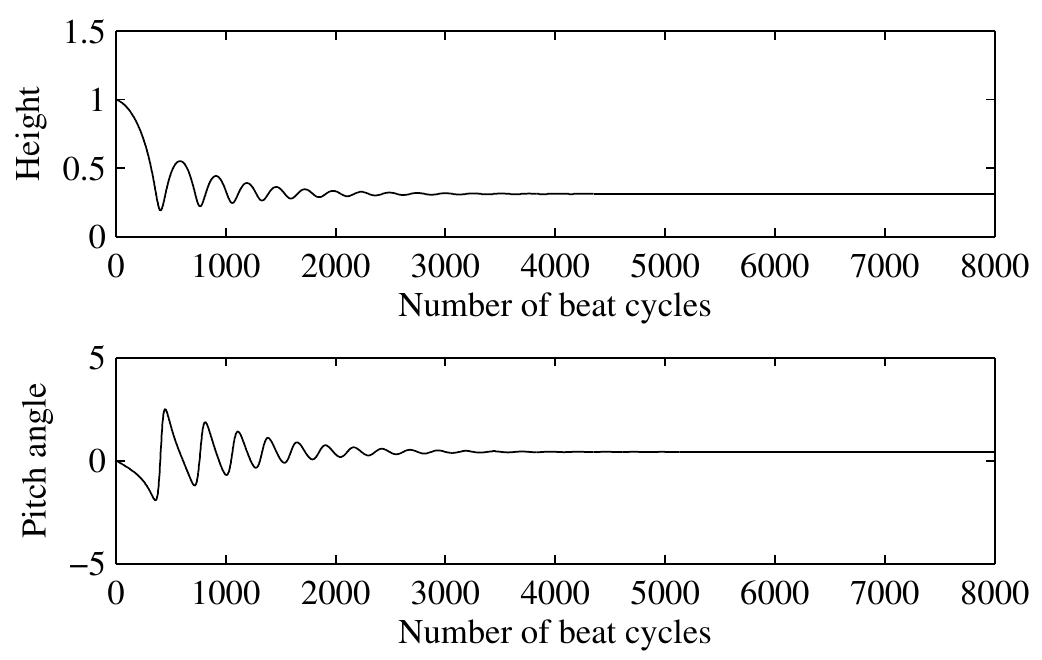}
}
\caption{(a) Schematic showing the definitions of height above the surface, measured to the head/flagellum junction, and pitch angle, measured from a plane parallel to the surface. (b) Height and pitch angle trajectories, from a simulation with wavenumber $k=2\pi$. (c) Height and pitch angle trajectories from a simulation with wavenumber $k=3\pi$.}\label{trajectories}
\end{figure}

Figure~\ref{trajectories}(a) shows the definition of height, measured from the surface to the head/ flagellum junction, and pitch angle, measured so that a positive value corresponds to pitching away from the surface. Simulation results for height and pitch angle with wavenumbers $k=2\pi$ and $k=3\pi$ are shown in Figure~\ref{trajectories}. These results are based on those presented in \cite{smithjfm}, which were the first long timescale simulations of sperm behaviour near surfaces, although the results we present here are for a greater number of timesteps. 

These results show that accumulation can occur purely through fluid dynamic effects: Figure~\ref{trajectories}(c) shows a relatively rapid convergence of a cell with wavenumber $k=3\pi$ to a specific height above the surface---around $x_3=0.31$. Figure~\ref{trajectories}(b) shows the behaviour of a cell with wavenumber $k=2\pi$. While it does not converge, it does suggest that after an initial period of swimming away, the cell will eventually return, swimming towards the surface. It is unclear whether this could ever result in swimming at a stable height, as for $k=3\pi$, and indeed whether the oscillations will grow or decay in amplitude. A simulation period of 8000 timesteps corresponds to an experimental timescale of 400--800 seconds, depending on the beat frequency. 

No specialised `bias' in the flagellar beat towards the surface is required to produce the accumulation behaviour, provided the wavenumber is sufficiently large. As observed experimentally by Winet et al.\ \cite{winet84}, cells eventually swim at a finite constant distance from the surface, as opposed to necessarily swimming against it. The surface does not simply attract the cell; it causes alternate pitching towards and away from the surface that steers the cell to this finite distance. Finally, perhaps counterintuitively, the `equilibrium state' of a cell swimming parallel to a surface involves a slight angle of inclination away from the surface.
In the following sections we investigate further the fluid mechanics of this behaviour.

\begin{figure}[H]
$
\begin{array}{ll}
\mbox{(a)} & \mbox{(b)} \\
  \raisebox{.83cm}{
   \includegraphics{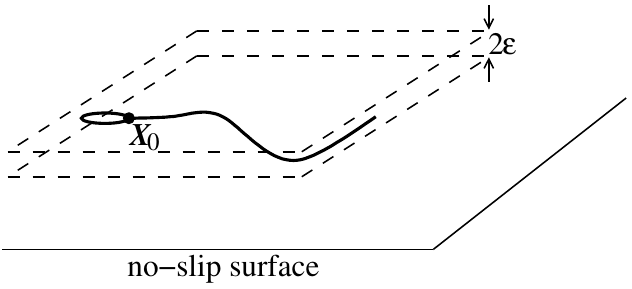}
   }
 & \scalebox{0.83}{
\includegraphics{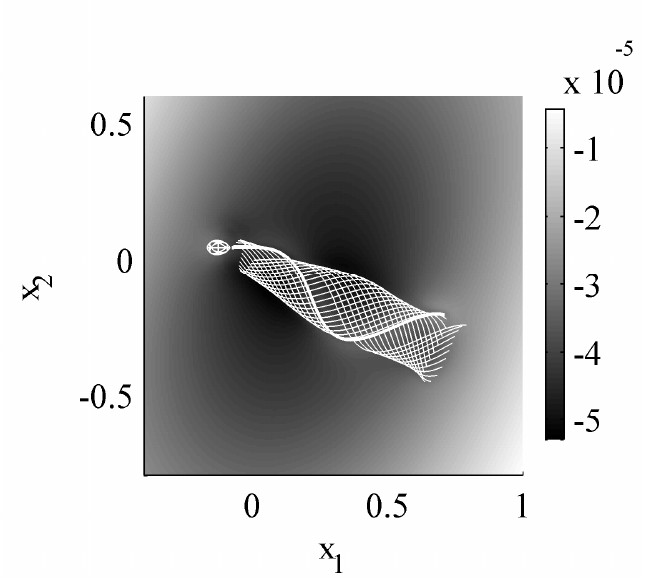}
}
\end{array}
$
\\
(c) \\
\includegraphics{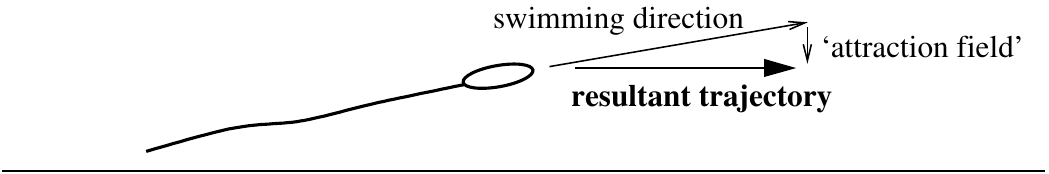}
\caption{(a) Schematic showing the planes parallel to the no-slip surface used for the calculation of the mean flow field around the cell.(b) Density plot showing the vertical component of the flow field $u_3$, averaged over one beat cycle, and averaged between the planes $Z_0-\varepsilon$ and $Z_0+\varepsilon$, where $\varepsilon=0.02$. (c) The equilibrium parallel-swimming behaviour of the cell involves a balance of inclined swimming away from the surface, and the attraction flow field shown in (b).}\label{flowfield}
\end{figure}

\section{The attraction field and pitching behaviour}
In order to gain insight into this effect, we calculate the time-averaged flow field around the cell, in particular the vertical component $\bar{u}_3$ that may push the cell towards or away from the surface. The flow fields evaluated at any finite distance above and below the cell show almost equal and opposite behaviour, so instead we calculate the average $\frac{1}{2}\left(\bar{\bs{u}}(\bs{X}_0+\varepsilon \bs{e}_3)+\bar{\bs{u}}(\bs{X}_0-\varepsilon \bs{e}_3)\right)$, as shown in Figure~\ref{flowfield}(a). 
In an infinite fluid, this average would be zero, however the image singularities of $\bs{\mathsf{B}}$ (equation~\eqref{blimage}) generate a non-zero vertical component, the dominant image singularity being the Stokes-dipole.

An example result for $\bs{X}_0=(0,0,1)$ and $k=3\pi$ is shown in Figure~\ref{flowfield}(b), showing that the cell generates an `attraction field', i.e. a vertical fluid velocity that pushes it towards the surface. This occurs for wavenumbers $k=2\pi$ and $k=3\pi$, and also occurs very close to ($x_3=0.1$) and further from ($x_3=1.0$) the surface. In accumulating cells, the eventual equilibrium trajectory occurs due to a balance of the attraction field, drawing the cell towards the surface, and a progressive component away from the surface, occurring because there is a small pitch angle away from the surface---see for example Figure~\ref{trajectories}(c), where the eventual pitch angle converges to $0.42^{\circ}$ while the height converges to $x_3=0.31$. This mechanism is summarised in Figure~\ref{flowfield}(c).

\section{The propulsive force distribution on the flagellum}
Cisneros et al.\ \cite{cisneros07}, using the related method of regularised Stokeslets, modelled the swimming of a bacterium near to a wall, representing the head as a sphere and the flagellum as a propulsive stick, with constant propulsive force distributed along its length. Surface accumulation behaviour was interpreted from the flow field generated around the swimmer, and also the existence of a resultant force predicted to act on a swimmer constrained to move parallel to a surface. In our simulations of sperm swimming we take a different approach: the propulsive force is calculated directly from the fluid/flagellum interaction, and cells are predicted to swim parallel to the surface while remaining free from any resultant force.

\begin{figure}[H]
$
\begin{array}{ll}
\mbox{(a), }k=2\pi & \mbox{(b), }k=3\pi \\
\scalebox{0.83}{
\includegraphics{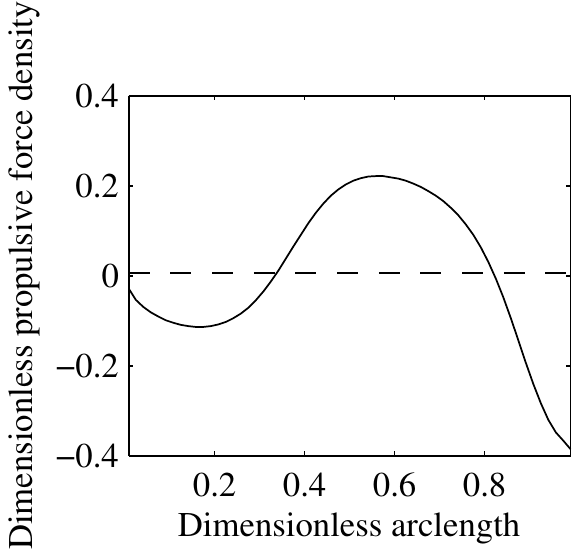}
} & 
\scalebox{0.83}{
\includegraphics{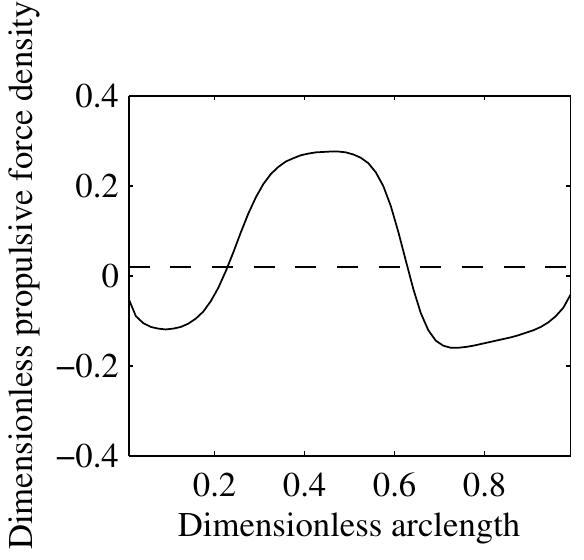}
}
\end{array}
$
\caption{Flagellar force distribution, averaged over one beat cycle for (a) $k=2\pi$ and (b) $k=3\pi$. The dotted line denotes the mean force distribution along the flagellum, showing that most of the propulsive effect generated by the mid region of the flagellum is balanced by nearly-equal and opposite drag occurring at the ends, as opposed to the head of the cell.}\label{meanF}
\end{figure}

While the attraction field appears to occur regardless of the wavenumber and distance from the surface, the pitching behaviour is very sensitive to both of these parameters, and it is this pitching effect, rather than the attraction field, that is responsible for the behaviour shown in Figure~\ref{trajectories}(b-c). Wavenumber affects the flow field through changes to the force distribution on the flagellum. In order to quantify how this changes with wavenumber, we calculate the time-averaged force distribution resolved in the direction of propulsion. Graphs of this quantity for $k=2\pi$ and $k=3\pi$ are shown in Figure~\ref{meanF}: both of the time-averaged force distributions differ significantly from the constant function used in the bacterial model. The average propulsive force along the flagellum, which balances the drag on the head and causes progression, is a fraction of the peak magnitude of the force along the flagellum: for a swimming cell such as a sperm with a relatively small cell body, the vast majority of drag occurs along certain portions of the flagellum, rather than the head. 

A fundamental theory of surface accumulation must take into account the complexity of the flagellar force distribution. Indeed, the relatively small change between $k=2\pi$ and $k=3\pi$ results in very significantly different behaviour (Figure~\ref{trajectories}). It should be noted that the bacterial flagellum differs substantially in structure and waveform from the sperm flagellum, and the force distribution for this cell type is likely to differ from our results in Figure~\ref{meanF}---and may more closely resemble a constant function.

\begin{figure}[H]
$
\begin{array}{ll}
 \mbox{(a)}                          
& 
 \mbox{(b)}
\\
 \raisebox{0.5cm}{
 $
 \begin{array}{c}
   \includegraphics{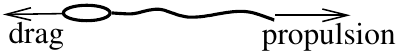}  
   \\
   \vspace{0.2in}
   \includegraphics{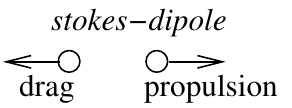}
 \end{array}
 $
 }
&
 \begin{array}{c}
 \includegraphics{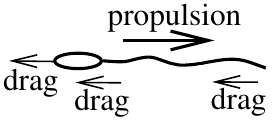}   
  \\
 \vspace{0.2in}
 \includegraphics{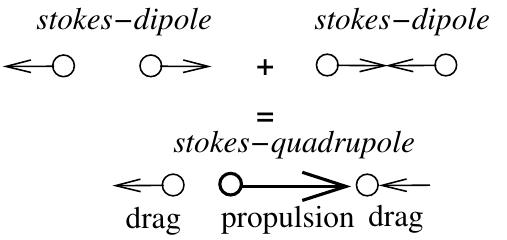}
 \end{array}
\end{array}
$
\caption{Two simplified models for the force exerted on the fluid by a sperm: (a) Stokes-dipole, (b) Stokes-quadrupole, the latter being inspired by the force distribution shown in Figure~\ref{meanF}(c).}\label{quadrupole}
\end{figure}

The simplest model of the fluid flow produced by a force-free swimmer is a stresslet, i.e. a Stokes-dipole that approximates a propulsive force due to the flagellum and a drag force due to the head acting very close together. This is based on the following simplified model: the drag on the head pushes fluid in front of the cell in the direction of swimming, while the propulsive effect of the flagellum pushes fluid rear of the cell in the opposite direction. Analysis of the propulsive force distribution Figure~\ref{meanF} shows that the situation is more complex for sperm motility. Figure~\ref{meanF} leads us to examine the hypothesis that the flow may be modelled more accurately as a Stokes-quadrupole $(\partial^2/\partial x_1^2) S_{j1}$, at least for $k=3\pi$, as shown in Figure~\ref{quadrupole}. This singularity has far-field decay $O(r^{-3})$, compared with the $O(r^{-2})$ behaviour of the Stokes-dipole. From our simulations, the decay of the velocity field around a sperm in an infinite fluid is approximately $O(r^{-2.4})$, intermediate between the Stokes-dipole and Stokes-quadrupole. When the no-slip surface is present, the decay of the velocity field is even more rapid, being approximately $O(r^{-3.8})$, as may be expected from presence of the additional image systems.

\section{The effect of the cell `head'}
\begin{figure}[H]
(a)\\
\scalebox{0.83}{
\includegraphics{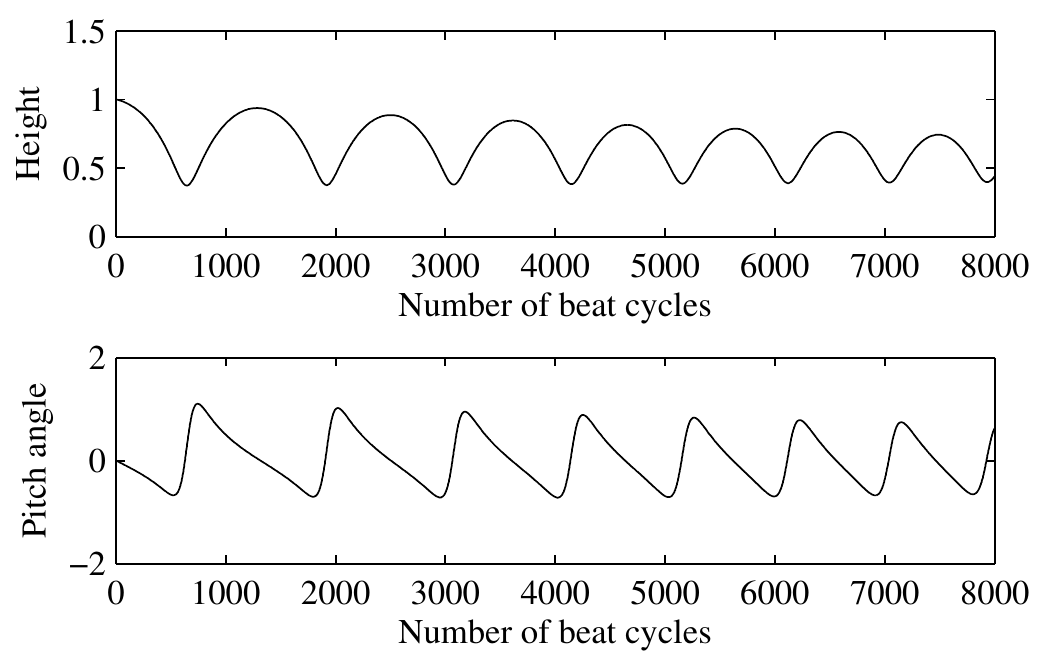}
}\\
(b)\\
\scalebox{0.83}{
\includegraphics{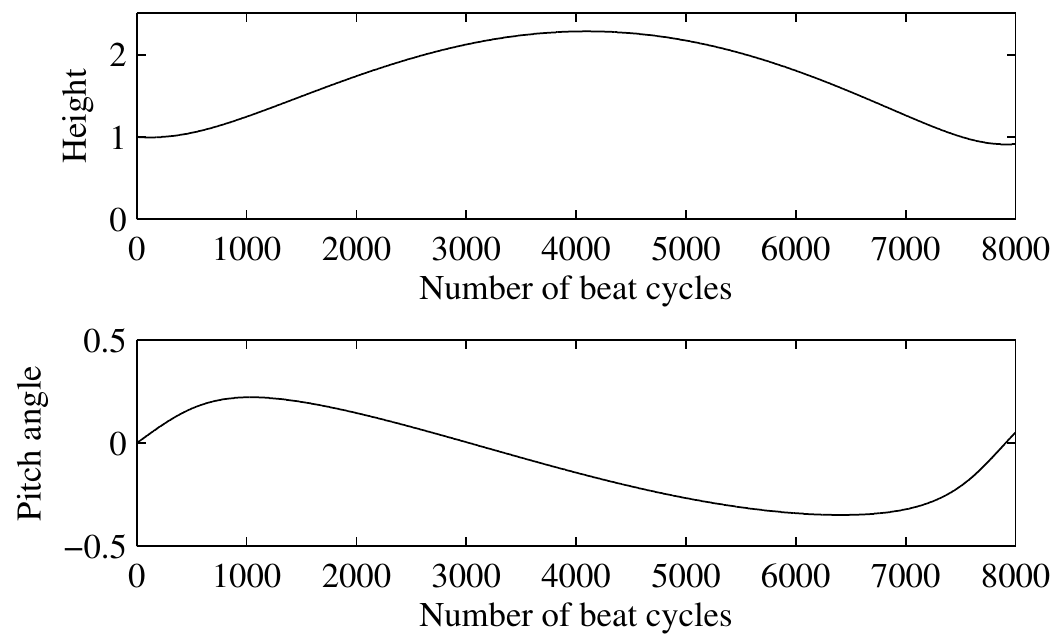}
}
\caption{Simulation results showing the effect of the head proportions on accumulation behaviour. (a) A cell with head dimensions half that of a normal cell, giving 1/8th the volume. This cell is predicted to accumulate, but the oscillations decay much more slowly than in the normal case. (b) A `pinhead' cell with head dimensions one tenth that of a normal cell, giving 1/1000th the volume. This cell is not predicted to accumulate over the timescale simulated of 4000 beat cycles, but nevertheless does exhibit an eventual pitching towards the surface.}\label{pinhead}
\end{figure}

Figure~\ref{meanF} suggests that the drag on the head, which is equal and opposite to the mean of the propulsive force on the flagellum, is very small in comparison with the peak flagellar force. This leads us to test whether the head has any significant effect on the accumulation behaviour. We perform simulations of a cell with half the usual head dimensions, resulting in $1/8$th the volume, and also a `pinhead' cell---as sometimes observed experimentally---with volume 1/1000th that of the normal head. The results are shown in Figure~\ref{pinhead}---a reduction in head dimensions results in much slower decay of the oscillations, while a pinhead cell does not even approach an equilibrium trajectory over 8000 beat cycles. Despite the fact that the head makes a relatively minor contribution to the drag in comparison with regions of the flagellum, it is nevertheless crucial to the surface accumulation effect. The reasons for this are unclear, but emphasise the subtlety of the fluid dynamic interaction.

\section{Conclusions}
Despite the apparent simplicity of the problem---modelling a single cell swimming in zero Reynolds number flow---surface accumulation is a relatively subtle problem and has evaded a simple explanation since the 1960s. Fluid dynamic simulations give a number of insights into the mechanism for this behaviour, and the roles of the head and flagellum force distributions. To summarise:
\begin{itemize}
\item For sufficiently large wavenumber, sperm have a `preferred' trajectory, in which they swim near and parallel to a surface, with zero total force.
\item This occurs through a combination of pitching behaviour, which `steers' the cell to an equilibrium height and a relatively weak attraction field, that balances the eventual slight inclination of the cell away from the surface.
\item Any theory of sperm accumulation must take into account the shape of the propulsive force distribution, and the fact that the sperm does not behave as a simple Stokes-dipole drag/propulsion combination.
\item The sperm head, while contributing a relatively small fraction of the drag, has very signficant effects on accumulation, in particular causing more rapid convergence of the cell trajectory.
\item In the real physiological system, sperm undergo transient spontaneous changes in swimming behaviour. This may be responsible for the fact that observations (for example, Figure 1) rarely show every single cell swimming stably at a fixed distance from the surface, as might be predicted from simulation studies. Furthermore this is an important difference between sperm and other swimming cells such as bacteria.
\end{itemize}
A `simple' explanation of this long-standing problem is surprisingly elusive. Moreover, similarities and differences between sperm and bacterial behaviour near surfaces require further investigation.

\section{Acknowledgements}
The authors thank Dr Jackson Kirkman-Brown of the Centre for Human Reproductive Science, Birmingham Women's NHS Foundation Trust for the unpublished experimental data shown in Figure 1, and for valuable discussions and guidance on the enigmatic world of sperm function and fertility. The authors also thank their colleagues Dr Eamonn Gaffney, Mr Henry Shum and Mr Hermes Gad\^{e}lha of the Mathematical Institute, University of Oxford for valuable discussions. The authors acknowledge comments from Dr Martin Bees, Department of Mathematics, University of Glasgow on the flow fields produced by bacteria, and Professor Tim Pedley, DAMTP, University of Cambridge, on the `fundamentals' of surface accumulation of sperm, which led to some of the investigations undertaken in this paper. David J.\ Smith was funded by a Medical Research Council Computational Biology Fellowship, grant number G0600178.

This is a preprint of a paper published in The Mathematical Scientist vol. 34 no. 2, pp. 74--87 (2009).

\end{document}